\documentclass[10pt]{article}

\usepackage{amsmath}
\usepackage{amssymb}

\usepackage{graphicx}
\let\oldbibliography\thebibliography
\renewcommand{\thebibliography}[1]{%
  \oldbibliography{#1}%
  \setlength{\itemsep}{0pt}%
}

\usepackage{color} 

\usepackage[labelfont=bf,labelsep=period,justification=raggedright]{caption}

\date{}

\begin{document}

\begin{flushleft}
{\Large
\textbf{MODULAR: Software for the Autonomous Computation of Modularity in Large Network Sets}
}

$~$\\
Flavia Maria Darcie Marquitti$^{1,\ast}$, 
Paulo Roberto Guimaraes Jr.$^{1}$, 
Mathias Mistretta Pires$^{1}$,
Luiz Fernando Bittencourt$^{2}$
\\
\textbf{1 Departamento de Ecologia, Universidade de S\~ao Paulo, S\~ao Paulo, SP 05508-900, Brazil}
\\
\textbf{2 Instituto de Computa\c{c}\~ao, Universidade Estadual de Campinas, Av. Albert Einstein, 1251, Campinas, SP 13083-852, Brazil}
\\
$\ast$ E-mail: flamarquitti@gmail.com\\
\end{flushleft}

\begin{abstract}
Ecological systems can be seen as networks of interactions between individual, species, or habitat patches. A key feature of many ecological networks is their organization into modules, which are subsets of elements that are more connected to each other than to the other elements in the network. We introduce MODULAR to perform rapid and autonomous calculation of modularity in sets of networks. MODULAR reads a set of files with matrices or edge lists that represent unipartite or bipartite networks, and identify modules using two different modularity metrics that have been previously used in studies of ecological networks. To find the network partition that maximizes modularity, the software offers five optimization methods to the user. We also included two of the most common null models that are used in studies of ecological networks to verify how the modularity found by the maximization of each metric differs from a theoretical benchmark.
\end{abstract}

\section*{Introduction}
Ecological systems can be seen as networks in which the elements, such as habitat patches within a landscape or species within communities, are represented as nodes and patch connectivity or species interactions are depicted as edges that connect the nodes \cite{dunne2006, urban2009}. The way such connections are organized affects system dynamics \cite{stouffer2011}, and thus how the system will respond to changes, such as species loss \cite{dunne2002} or changes in the ecological connectivity among patches \cite{cumming2010}. Several descriptors have been used to characterize the organization of networks, i.e., the network topology \cite{dunne2006, carstensen2012}. A recurrent pattern in ecological networks is modularity, also termed compartmentalization, clustering, or community structure \cite{boccaletti2006}. Modules are cohesive groups of highly connected nodes that are loosely connected to other nodes in the network \cite{newman2004, olesen2007}.

In ecological networks, modularity will emerge if there are certain groups of individuals, species, or habitat patches that that show more interactions among each other than with other groups within the network \cite{guimacurrent, tinker2012, dale2010, carstensen2012}. This type of modular organization has been found in networks describing different ecological systems, such as resource use by animal populations \cite{araujo2008}, food webs \cite{krause2003, allesina2009}, mutualistic interactions between plant species and their pollinators \cite{olesen2007}, antagonistic interactions between parasites and their hosts and between plants and their herbivores \cite{fortuna2010, pires2012}, and the spatial connectivity of metapopulations \cite{bodin2007, dale2010}. Because the degree of modularity determines how dense the connection between different groups of elements in an ecological system is, systems that largely differ in the degree of modularity often differ in their ecological and evolutionary dynamics \cite{guimacurrent, olesen2007, thebault2010, cumming2010, hagen2012}.	

Propelled by the relevance of modularity in the dynamics of ecological systems, the detection of modularity became an important aspect of studies that analyze the organization and conservation of ecological networks \cite{cumming2010}. Modularity can be measured by different metrics, and one of the most popular modularity metrics in the study of ecological networks is the $Q$ metric (also referred to as \textit{M})\cite{newman2004, guimera2005, olesen2007, fortuna2010}. For a given partition of a network into modules, \textit{Q} measures the proportion of edges that connect the nodes within the same module. Because finding the maximal modularity (maximal \textit{Q}) in a network is an NP-hard problem \cite{ruan2007}, there is no known algorithm to find the maximum \textit{Q} in polynomial time. Thus, it is necessary to use sub-optimal optimization approaches to perform this calculation. Heuristic algorithms can thus be used to approximate a network partitioning with the largest number of edges within the modules and the lowest number of edges between modules to maximize the modularity index (see details of the metrics in Supplementary material Appendix 1). Although module identification is scale-dependent, optimization algorithms can be used to test module consistency across multiple scales, testing the effects of resolution on module detection \cite{fortunato2007}.

Finding the partition with the highest modularity in a large network is often time consuming. Moreover, the analysis of large sets of data and the subsequent testing of the results against theoretical benchmarks that are represented by ensembles formed by thousands of replicates is a common procedure in biology \cite{gotelli1996, gotelli2001}. In this sense, a major constraint in the analysis of the modularity of ecological networks is the lack of programs that allow fast and autonomous computation of modularity for researchers who are not familiar with programming. Here, we introduce the software MODULAR for the computation of the modularity and the identification of modules in multiple complex networks. Many algorithms have been proposed for finding the partition that maximizes the value of $Q$, and some of these are publicly available. However, to the best of our knowledge, there is no software available that allows multiple uses of different metrics and optimization algorithms in a user-friendly way that accelerates the workflow of the ecologist. MODULAR was developed to allow one to automatically compute the modularity of several input files and to allow one to choose the optimization algorithm that best matches the user's needs. 

\section*{MODULAR features}
MODULAR was developed in \textit{C} language and uses features from the igraph-0.6 library \cite{igraph} and the GNU Scientific Library (GSL) \cite{GSL} (see details in Supplementary material Appendix 2). MODULAR was designed to facilitate and speed up the detection of modules in multiple networks through modularity maximization. To achieve this task, the maximization of modularity is automatically performed for a set of input files containing representations of bipartite networks, such as those depicting species occurrence across islands \cite{carstensen2012}, or unipartite networks, such as spatial networks describing habitat connectivity for a given species \cite{dale2010}. 

When running MODULAR with a representation of unipartite networks, only the \textit{Q} metric is available \cite{newman2004}. If the input data files represent bipartite networks, the user can choose between two different modularity metrics: Newman and Girvan's - Q \cite{newman2004} or Barber's modularities - $Q_B$ \cite{barber2007}, which is a modification of the \textit{Q} metric for bipartite networks (see details of metrics in Supplementary material Appendix 1). The two options are available for bipartite networks because unipartite indexes have also been used in the ecological literature for bipartite networks (e.g., Olesen \textit{et al.}, 2007; Fortuna \textit{et al.}, 2010 Carstensen \textit{et al.}, 2012).

If the user chooses the traditional \textit{Q} metric, there are five optimization algorithms that can be used to perform the search for the partition of the network into modules that maximizes the modularity index: (i) fast greedy (FG) \cite{clauset2004, wakita2007}, (ii) simulated annealing (SA) \cite{guimera2005}, (iii) spectral partitioning (SP) \cite{newman2006}, (iv) a hybrid of simulated annealing and spectral partitioning (Hyb-SP), and (v) a hybrid of simulated annealing and fast greedy (Hyb-FG). The optimization algorithms differ in the method used to search the network partition that maximizes the modularity measurement (see details of MODULAR functioning and optimization algorithms in Supplementary material Appendix 3). Since the running time can vary considerably, the choice of the optimization algorithm becomes particularly important.

To verify whether the modularity found by the maximization of each metric significantly differs from a theoretical benchmark, the user has the option of running two different null models and to specify how many replicates each null model will generate. We included unipartite and bipartite versions of two of the most common null models that are used in studies of ecological networks: (i) the Erd\"{o}s-R\'{e}nyi model \cite{erdos1959} and (ii) the ``null model 2'' \cite{bascompte2003} (see details of the null models in Supplementary material Appendix 4). Because MODULAR can utilize large sets of input data, the user can also test the modularity of the networks generated by other null models by using those as input data.

MODULAR is an open source software program licensed under the GNU General Public License version 3. It can be downloaded from http://sourceforge.net/projects/programmodular. As future directions, we are planning to add new algorithms and optimization methods for the calculation of modularity. In addition, new metrics that analyze modularity at the node level can be added to MODULAR.

\section*{Acknowledgments}
We thank the igraph developers for promptly solving all of the reported issues in the external interactions of the library. We would also like to thank all of our colleagues who encouraged us to persist in developing this software and who tested it in the earlier stages. We are very thankful to Pedro Jordano for his comments and suggestions. We thank FAPESP for financial support.

\bibliography{BiblioExample}

\newpage
{\centering
\section*{Supplementary material}}

\section*{Appendix 1: Metrics}
The metrics used in MODULAR are the Newman and Girvan (2004) \cite{newman2004} metric and its modified version for bipartite networks, which was proposed by Barber (2007) \cite{barber2007}. The first metric is calculated as follows:

\begin{equation}
Q = \sum_{i=1}^{N_{_{M}}} \left[ \frac{E_i}{E} - \left(\frac{k_i}{2E}\right)^2 \right],
\end{equation}
where $N_{_{M}}$ is the number of modules, $E_i$ is the number of links in module $i$, $E$ is the number of links in the complete network, and $k_i$ is the sum of the degrees of the nodes within module $i$. Thus, a good partition has many links inside the modules and as few links as possible between the modules.

The bipartite version of the metric is calculated as follows:

\begin{equation}
Q_B = \sum_{i=1}^{N_{_{M}}} \left[\frac{E_i}{E} - \left(\frac{{{k_i}^C}\cdot{{k_i}^R}}{E^2}\right) \right],
\end{equation}
where ${k_i}^C$ is the sum of the degrees of the nodes within module $i$ that belong to set $C$ and ${k_i}^R$ is the sum of the degrees of the nodes within module $i$ that belong to set $R$.

\section*{Appendix 2: Language, Libraries, and Algorithms}
We implemented the MODULAR software in \emph{C}. Our code depends on two libraries: igraph and the GNU Scientific Library (GSL). The igraph library provides data structures for graph representation, manipulation functions and functions for community structure detection. The GSL library is mainly used to generate the random numbers that are needed by the optimization methods. By implementing a set of functions utilizing those two libraries, we were able to develop an easy-to-use software program for the detection of modules with maximum modularity. The igraph library utilized is the $0.6$ version, which supports UCINET's DL file format. We used the \textit{igraph\_read\_graph\_dl} function to read the network into a graph data structure. In addition, we used igraph data structures called \textit{membership vectors} to represent the modules of the network. The SA \cite{simulatedAnnealing} method was implemented using a data structure from GSL that contains a number of running parameters, such as the initial temperature and its damping factor. The \textit{fast greedy} \cite{clauset2004} and the \textit{spectral partitioning} \cite{newman2006} methods are functions called from the igraph library. These two methods are faster than simulated annealing, but their searches are based on local decisions and thus eventually lead to lower modularities than those found with SA. In MODULAR, we combined the speed of these two methods to find a reasonable solution with the broader search that is provided by the simulated annealing method by creating two hybrid methods. The first hybrid approach (\textit{Hyb-SP}) first runs the SP method, and its module configuration output is utilized as the input for the SA method using a reduced initial temperature. The same technique is applied in the second hybrid method, except that the FG method is used instead of the SP method (\textit{Hyb-FG}).

\section*{Appendix 3: MODULAR Functioning}
\begin{figure}[!ht]
\begin{center}
\includegraphics[width=0.9\textwidth]{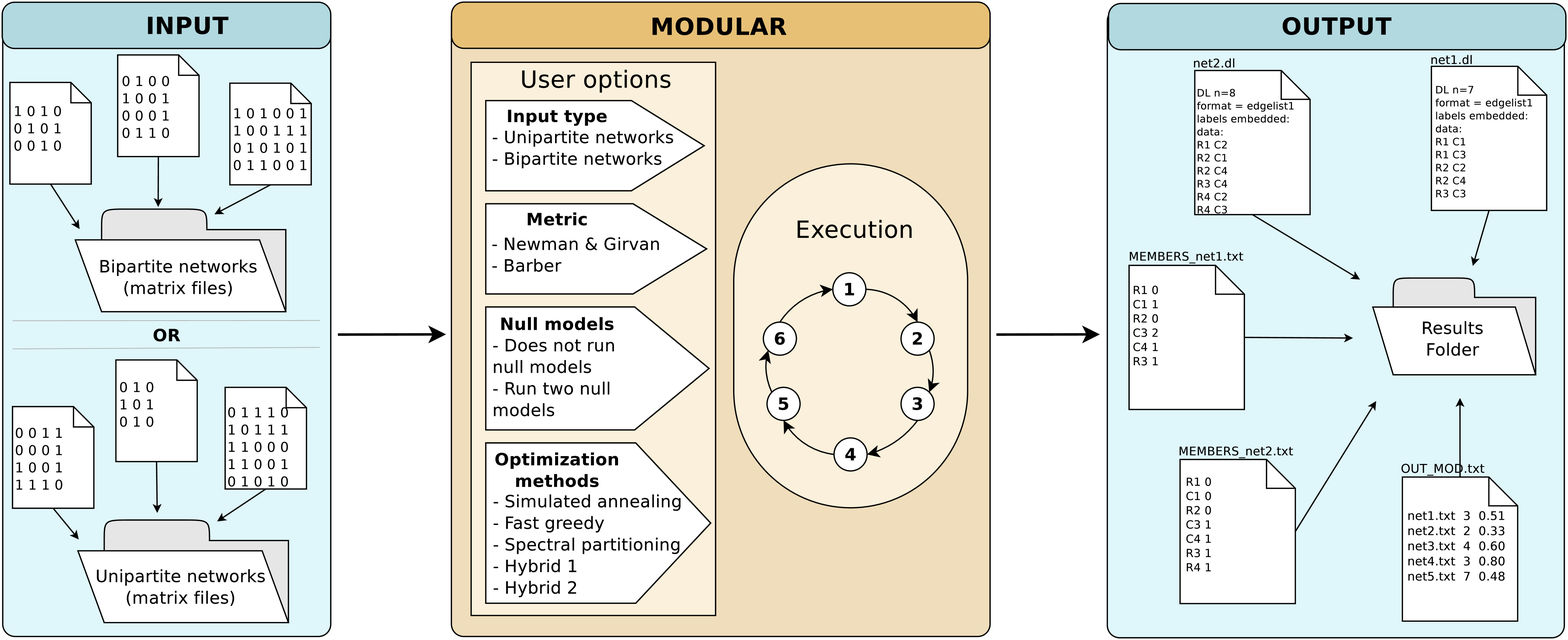}
\end{center}
\caption{
\textbf{Design of the MODULAR software}. The program design can be divided into three steps. The first step is the \textit{input} file preparation. The user may choose between representations of bipartite or unipartite networks. The second step includes the running of \textbf{MODULAR}. After the user answers some questions, MODULAR will execute six steps (see the execution explanation in the text). Finally, MODULAR will produce \textit{output} files with the results. 
 }
 \label{fig:design}
\end{figure}

The basic MODULAR functioning is represented in Figure A1\ref{fig:design}. The set of input files can include two kinds of files: networks represented as matrices in text files (extension .txt) or list of interactions, such as those used by the UCINET program (extension .dl) \cite{UCINET}. Both kinds of input files can contain representations of \textit{bipartite networks} or \textit{unipartite networks}. If the input data are text files containing matrices, each \textit{.txt} file is a binary matrix in which the columns are separated by tabulation or a space. The user must not include any row or column labels in the input file. The representation of unipartite networks should contain square matrices in which the rows \textit{i} and columns \textit{i} depict the same element in the \emph{N$\times$N} matrix. These matrices are symmetrical, and the program will read only the input data from the upper triangular part. 

The current version of the program is not intended to be used with directed networks. If the matrices in the text files represent bipartite \emph{N$\times$M} networks, the rows and columns represent different sets of elements, and edges exist only between elements of the two different sets. For example, the rows can represent islands and the columns can represent species in a network describing the species co-ocurrence across islands \cite{carstensen2012}. In contrast, if the input data are UCINET files, the representation of unipartite networks should follow the UCINET file formats with the \emph{format} field set to contain the format type \textit{edgelist1} followed by an edge list of the node pairs. The representation of bipartite networks should set the \emph{format} field to \textit{edgelist2}, followed by an edge list of the node pairs. The first node of the pair is an element of one set, and the second node is an element of the other set. Therefore, the first option that MODULAR gives the user is the selection of different types of input files that represent bipartite and unipartite networks.

The second option given to the user regards the modularity metric: Newman and Girvan's - Q \cite{newman2004} or Barber's modularities - $Q_B$ \cite{barber2007} (see the Appendix 1 for details). If the user chooses the traditional \textit{Q} metric, there are five optimization algorithms that can be used: (i) fast greedy (FG) \cite{clauset2004, wakita2007}, (ii) simulated annealing (SA) \cite{guimera2005}, (iii) spectral partitioning (SP) \cite{newman2006}, (iv) a hybrid of simulated annealing and spectral partitioning (Hyb-SP), and (v) a hybrid of simulated annealing and fast greedy (Hyb-FG). The optimization algorithms differ in the method used to search the network partition that maximizes the modularity measurement (objective function). 

The FG method searches the maximum modularity by only accepting partitions that exhibit a higher modularity than the previous partition. It is a greedy optimization method. It was proposed by Clauset \textit{et. al} (2004) for the analysis of modularity and improved by Wakita \& Tsurumi (2007). The method is summarized below:\\

\textbf{FG optimization:}
\begin{itemize}
\item[] 1) Compute the modularity of the current partition of the network.
\item[] 2) Join the communities and accept the new partition that provides the highest modularity.
\item[] 3) Stop when no partition joining can increment the modularity.
\end{itemize}

The SA method searches the maximum modularity while trying to avoid getting stuck at local maxima \cite{simulatedAnnealing}. This time-consuming method first partitions the network into modules with one node each, thus generating an initial configuration of \emph{N} modules of size 1. The method then randomly moves nodes between modules and computes the resulting modularity of the new module configuration. If the new modularity is higher than that obtained with the previous module configuration, the algorithm takes the new module configuration as the current solution. However, if the new modularity is lower than that obtained with the previous module configuration, the algorithm accepts it as the new solution with a given probability \cite{metropolis1953}, thus simulating a system of particles in which the energy levels follow a Boltzmann distribution. Using this technique, the algorithm avoids getting trapped at local maxima by moving through sub-optimal solutions and potentially converging to a maximal point \cite{simulatedAnnealing}. This avoidance is possible because of the \emph{temperature} parameter, which is directly linked to the probability of accepting new solutions. The temperature decreases gradually during the process, thereby decreasing the probability to accept worse solutions (lower modularity values). The steps are repeated until the SA temperature reaches a threshold or the algorithm undergoes a predetermined number of iterations without any changes in the solution. If the user chooses to maximize Barber's metric, the simulated annealing method is the only search method available in the current version of the program. The method is summarized below:\\ 

\textbf{SA optimization:}
\begin{itemize}
\item[] 1) Compute the modularity in the current partition of the network.
\item[]  2) Generate a new module partitioning by randomly exchanging nodes and merging/splitting modules, compute the modules, and determine the modularity.
\item[] 3) Accept the new partition if the modularity increases. If the modularity decreases, accept the new partition with a given probability based on the difference between the new and the previous modularity values and according to the chosen temperature parameter used to run the SA.
\item[] 4) Stop if the solution does not change after a predetermined number of iterations or when the minimum temperature chosen to run the SA is reached.
\end{itemize}

The SP method is based on matrix spectra. This fast method divides the network into sequential groups according to the entries in the leading eigenvector \cite{clauset2004}.\\

\textbf{SP optimization:}
\begin{itemize}
\item[] 1) Establish the modularity matrix (matrix B, see Newman 2006).
\item[] 2) Divide the network into two groups according to the entries in the leading eigenvector. 
\item[] 3) Repeat the process until none of the sub-matrices have a positive eigenvalue.
\end{itemize}

In MODULAR, we introduced two additional hybrid methods that combine the speed of the SP and FG approaches with the larger search space of simulated annealing. In the hybrid approaches, the SA input is the output of the faster method (SP or FG). Consequently, the optimization can achieve better results. After these steps, the user can choose to verify the modularity found by the maximization against two different null models.

Based on the options above, MODULAR iteratively executes up to six steps (``Execution'', in Figure A1) for each \textit{.txt} or \textit{.dl} file present in the matrices folder specified by the user:

For each \textit{.txt} or \textit{.dl} file in the specified folder, do:

\begin{itemize}
	\item[] 1) Read the matrix $A$ or the edge list from the input file according to the specified type (representation of a bipartite or unipartite network);
	\item[] 2) If the input file is a matrix, remove all null rows and null columns present in the matrix to generate a new matrix $A'$;
	\item[] 3) If the input file is a matrix, generate a \textit{.dl} UCINET \cite{UCINET} file with the network represented by matrix $A'$;
	\item[] 4) Read the \textit{.dl} file and load the network into the computer memory using data structures that can be manipulated by the \textit{igraph} library functions;
	\item[] 5) Maximize the modularity of the network by running the method selected by the user (SA, SP, FG, Hyb-SP, or Hyb-FG);
	\item[] 6) Generate the theoretical networks according to the null model (if requested by the user) and run step 5 for each generated network.
\end{itemize}

A set of output files is generated after these steps. There are three main output files. If the input files are named \textit{MatrixName.txt}, MODULAR generates a file with the list of interactions named \textit{MatrixName.dl} in the UCINET edgelist1 format. The first line indicates the number of nodes in the network. If MODULAR was used to analyze bipartite network representations, the rows will be named \emph{R} followed by a number (for example, \emph{R2} for the second row of the input file), and the columns will be named \emph{C} followed by a number (for example, \emph{C3} for the third column of the input file). If the data set is composed of unipartite network representations, the lines will be named \emph{L} followed by a number (for example, \emph{L2} for the second line of the input file). In both cases, the number refers to the order found in the input file.

A set of output files is generated after these steps. The main output is the \textit{OUT\_MOD.txt} file. This file contains, for each input file, the value of the modularity metric and the number of modules found by the metric and algorithm chosen. If the user chooses to run the null models, the \textit{OUT\_MOD.txt} file will also contain the proportion of null matrices with modularity higher than that in the input file. An output file named \textit{MEMBERS\_MatrixName.txt} is generated for each input file named \textit{MatrixName.txt}. This file has two columns: the first column has the labels of the matrix lines from the \textit{.dl} file and the second column indicates the module to which each line belongs. If null model analysis is performed, there will be two additional outputs: \textit{OUT\_1\_MatrixName.txt}, which includes the modularities of null model 1, and \textit{OUT\_2\_MatrixName.txt}, which includes the modularities of null model 2. Both files have as many lines as the number of null model runs chosen by the user.

\section*{Appendix 4: Null models}
The null models implemented in MODULAR are the Erd\~{o}s-R\'{e}nyi model \cite{erdos1959} and ``null model 2'' \cite{bascompte2003}. In the Erd\~{o}s-R\'{e}nyi model, each pair of nodes has the same probability of being connected by an edge. Thus, the probability of a pair $(i,j)$ to be linked is given by
 
\begin{equation}
P (i,j) =  \frac{E}{{R}\cdot{C}},
\end{equation}
where $E$ is the number of edges in the network. If the network is bipartite, $R$ and $C$ are the number of nodes in each set of the network. For unipartite networks, $R=C$ and the probability of a pair $(i,j)$ to be linked is given by

\begin{equation}
P (i,j) =  \frac{2E}{{R}\cdot{(R-1)}}.
\end{equation}
 
In ``null model 2'' \cite{bascompte2003}, the probability of a pair being connected by an edge is proportional to the number of edges that the nodes have. Thus, the probability of a pair $(i,j)$ to be linked is given by

\begin{equation}
P (i,j) =  \frac{1}{2} \left(\frac{k_{_{i\in R}}}{C} + \frac{k_{_{j\in C}}}{R} \right),
\end{equation}
where $k_i$ is the number of edges of node $i$ in the $R$ partition and $k_j$ is the number of edges of node $j$ in the $C$ partition. For adjacency matrices, $R=C$. 

For additional information on the potential uses of null models, please refer to the study published by Ulrich and Gotelli \cite{ulrich2007}.

\end{document}